\documentclass[osajnl,twocolumn,showpacs,superscriptaddress,10pt]{revtex4-1} %% use 11pt for Applied Optics

\usepackage{amsmath,amssymb,bm,graphicx}
\newcommand{\uline}[1]{{#1}}
\usepackage{color}
\usepackage{stmaryrd} % for \llbracket
\usepackage{siunitx} % si units

\usepackage{csquotes}   % for enquote

\definecolor{BrickRed}{RGB}{153,0,0}
\definecolor{OliveGreen}{RGB}{0,153,0}
\definecolor{DarkBlue}{RGB}{0,0,153}

\newcommand{\UCL}{Department of Electronic and Electrical Engineering, University College London, Torrington Place,\\London WC1E 7JE, UK}
\newcommand{\PhotonD}{Photon Design Ltd, 34 Leopold Street, Oxford OX4 1TW, UK}

\renewcommand{\emph}[1]{{\textit{#1} }}

\newcommand{\E}{{\bf E}}        % electric field
\newcommand{\D}{{\bf D}}        % electric displacement field
\renewcommand{\P}{{\bf P}}  % total source polarization
\newcommand{\p}{{\bf p}}        % single term of source polarization
\renewcommand{\j}{{\bf j}}
\newcommand{\J}{{\bf J}}        % source current

\newcommand{\te}{{\bf e}}
\newcommand{\tm}{{\bf h}}
\newcommand{\amp}{{\bf a}}
\newcommand{\Gf}{{\bf G}}   % Green's function
\newcommand{\heavi}{{H}}
\newcommand{\fmatrix}[1]{{\left\llbracket #1 \right\rrbracket}} % matrix of Fourier coefficients
\newcommand{\fvec}[1]{{\left[#1\right]}}    % vector of Fourier coefficients

\newcommand{\bfchi}{ {{\boldsymbol\chi}^{(2)}  } }

\newcommand{\ez}{{\bf e}_z}     % unit vector in z-direction
\newcommand{\x}{{\bf x}}

\newcommand{\K}{{\bf K}}
\renewcommand{\k}{{\bf k}}
\newcommand{\height}{{h}}

\newcommand{\Nb}{{{\bf N}_b}}   % background solution operator
\newcommand{\polNL}{{ {\bf P}^\text{NL}}}   % nonlinear polarization
\newcommand{\deps}{\Delta \varepsilon}
\newcommand{\rec}[1]{{{\bf R}\left(#1\right)}}  % reconstruction operator

\newcommand{\Niter}{N_\text{iter}}  % number of iterations

\newcommand{\figref}[1]{{Fig.~\ref{#1}}}
\newcommand{\tabref}[1]{{Table~\ref{#1}}}

\def\minus{
  \setbox0=\hbox{-}
  \vcenter{
    \hrule width\wd0 height \the\fontdimen8\textfont3
  }
}
\def\vspacer{
    \textcolor{white}{\!\!\!\huge|\!\!\!\!\!}
}
\SetSymbolFont{stmry}{bold}{U}{stmry}{m}{n}

\def\CaF2{CaF$_2$}

\def\am{Adv.\ Mater.\ }
\def\ao{Appl.\ Opt.\ }
\def\jlt{IEEE\ J.\ Lightwave\ Technol.\ }
\def\jqe{Quantum. Electron.\ }
\def\jo{J.\ Opt.\ }
\def\josaa{J.\ Opt.\ Soc.\ Am.\ A.\ }
\def\josab{J.\ Opt.\ Soc.\ Am.\ B.\ }
\def\nat{Nature\ }
\def\nl{Nano\ Lett.\ }
\def\nt{Nanotechnol.\ }
\def\oe{Opt.\ Express\ }
\def\ol{Opt.\ Lett.\ }
\def\pr{Phys.\ Rev.\ }
\def\prb{Phys.\ Rev.\ B\ }
\def\prl{Phys.\ Rev.\ Lett.\ }
\def\sci{Science\ }

\begin{document}

\title{Nonlinear generalized source method for modeling second-harmonic generation in diffraction gratings}

\author{Martin Weismann}\email{martin.weismann.12@ucl.ac.uk}    % Corresponding author's email:
\affiliation{\UCL}
\affiliation{\PhotonD}
\author{Dominic F. G. Gallagher}
\affiliation{\PhotonD}
\author{Nicolae C. Panoiu}
\affiliation{\UCL}

\begin{abstract}
We introduce a versatile numerical method for modeling light diffraction in periodically patterned
photonic structures containing quadratically nonlinear non-centrosymmetric optical materials. Our
approach extends the generalized source method to nonlinear optical interactions by incorporating
the contribution of nonlinear polarization sources to the diffracted field in the algorithm. We
derive the mathematical formalism underlying the numerical method and introduce the
Fourier-factorization suitable for nonlinear calculations. The numerical efficiency and runtime
characteristics of the method are investigated in a set of benchmark calculations: the results
corresponding to the fundamental frequency are compared to those obtained from a reference method
and the beneficial effects of the modified Fourier-factorization rule on the accuracy of the
nonlinear computations is demonstrated. In order to illustrate the capabilities of our method, we
employ it to demonstrate strong enhancement of second-harmonic generated in one- and
two-dimensional optical gratings resonantly coupled to a slab waveguide. Our method can be easily
extended to other types of nonlinear optical interactions by simply incorporating the
corresponding nonlinear polarization sources in the algorithm.
\end{abstract}


\maketitle %% required

%%%%%%%%%%%%%%%%%%%%%%%%%%%%%%%%%%%%%%%%%%%%%%%%%%%%%%%%%%%%%
\section{Introduction}
\label{sec:Introduction} Higher-harmonic generation is a fundamental nonlinear optical process
that has attracted intense interest ever since the beginnings of nonlinear optics
\cite{bp62pr,p63pr}. Generation of optical waves via nonlinear interaction provides an effective
approach to explore the properties of light matter interaction at fundamental level and, equally
important, has found key applications in many areas of science and technology. While most of the
early studies of higher-harmonic generation focused on wave interaction in homogeneous nonlinear
optical media, recent advances in nanofabrication techniques and the advent of metamaterials have
considerably broadened the experimental and theoretical framework in which these nonlinear optical
processes are explored. In particular, second-harmonic generation (SHG) from isotropic
\cite{fzp06nl,kew06s,nsh06prl} and chiral \cite{vsv10prl,vbc14am} metasurfaces, layered media
coupled to arrays of plasmonic particles \cite{cdh07nl}, and quantum engineered plasmonic
metasurfaces \cite{lta14n} has been observed.

A key prerequisite condition for achieving efficient SHG is that the wavevectors of the
interacting waves are phase-matched. One particularly efficient method to phase match optical
waves is by using diffraction gratings. In this approach, the residual wavevector imbalance is
canceled by the grating wavevector, $\kappa=2\pi/\Lambda$, where $\Lambda$ is the grating period.
Importantly, the wide range of geometrical and materials parameters characterizing a diffraction
grating allows one to employ such optical devices to achieve wavevector phase matching in a
multitude of optical wave configurations. More importantly in this context, strong optical field
enhancement and, consequently, a significantly more efficient nonlinear optical interaction can be
achieved by using metallic gratings. Because generally there is no explicit analytical solution to
the problem of diffraction by periodic structures, efficient and accurate numerical methods are
essential for the design of diffraction gratings with predefined spectral characteristics. Solving
this problem in the nonlinear case is an even more daunting task, chiefly because the larger
number of interacting waves and their intricate interplay significantly increases the complexity
of the problem.

The periodic nature of diffraction gratings renders frequency (wavevector) domain methods based on
the Floquet-Bloch theorem to be particularly suitable algorithms for describing such optical
structures. A common characteristic of these methods is the Fourier expansion of the optical field
in a suitably chosen basis of modes, the main unknowns to be found being the corresponding Fourier
coefficients.
Among the most used methods in this class one should mention the Fourier modal method (FMM),
also known as rigorous coupled-wave analysis (RCWA)
\cite{mgp95josaa,l97josaa,lm96josaa,srk07josaa,eb10oe},
the \textit{C}-method \cite{crm80jo,lcg99ao} for corrugated gratings, and Green's function
type of methods \cite{s87josab,ms05josaa,st12jqsrt}.
On the other hand, there is a scarcity of numerical methods for SHG and other nonlinear interactions
in periodic structures and the convergence and runtime behavior of the few that do exist are not
as well characterized. These methods are either problem specific algorithms \cite{pn94josab} or
extensions of linear methods to the nonlinear case, e.g. the FMM
\cite{ntf02josaa,btf07josab,prl10josab}, finite-difference time-domain method
\cite{lh04josab,rmj06jlt}, Green's function method \cite{cy02prb}, and multiple scattering matrix
method \cite{bp10prb,xz12oe}.

In this work we show how the linear generalized source method (GSM) \cite{st12jqsrt,st10jqe} can
be adapted to describe the SHG in diffraction gratings containing non-centrosymmetric materials.
We choose the linear GSM as a starting platform for our new method because it has proven to be an
efficient alternative to the FMM for the analysis of one- and two-dimensional (1D, 2D),
corrugated, periodic optical structures. Its optimal runtime is of order $N_o\log(N_o)$, where
$N_o$ is the total number of diffraction orders, as compared to $N_o^3$ for the conventional FMM.
The key ingredients needed to achieve this remarkable computational efficiency is the use of a
matrix-multiplication based iterative solver, namely the generalized minimum residual (GMRES)
method, to solve an underlying linear system of coupled integral equations and a fast
Toeplitz-matrix-vector multiplication performed by means of a fast-Fourier transform (FFT). In
addition, its implicit, Green's function type formulation of Maxwell equations allows for a natural
incorporation of nonlinear source terms, which is the main challenge in computing the SHG
response.

The rest of the paper is organized as follows: in Section~\ref{sec:Derivation} we introduce the
formulation of the nonlinear GSM, including the correct Fourier-factorization rule, and present
the proper discretization of the underlying equations that allows one to use a fast iterative
solver. Then, in Section~\ref{sec:Benchmark}, the convergence and runtime behavior of the method
are studied. The method is applied to optimize the radiated second-harmonic (SH) by a diffraction
grating coupled to a slab waveguide in Section~\ref{sec:Application} before final conclusions are
drawn in Section~\ref{sec:Conclusion}.

 %%%%%%%%%%%%%%%%%%%%%%%%%%%%%%%%%%%%%%%%%%%%%%%%%%%%%%%%%%%%%
\section{Derivation of the nonlinear generalized source method}
\label{sec:Derivation} We start the derivation of the nonlinear GSM from the governing equations
for SHG in the frequency domain and describe the general solution strategy. We then derive the
nonlinear GSM for 2D periodic structures following \cite{st12jqsrt} and extend its scope to the
SHG in optical gratings containing non-controsymmetric materials.

\subsection{Physical model for second-harmonic generation}
We consider a quadratically nonlinear optical medium and a time-harmonic electric field,
$\bm{\mathcal{E}}(\x,t)$, propagating in the medium at the fundamental frequency (FF),
$\omega_{0}$,
\begin{eqnarray}
   \bm{\mathcal{E}}(\x,t) = \E(\x)e^{i\omega_{0} t} + c. c.,
\end{eqnarray}
where $\x$ and $t$ are the position vector and time, respectively. Through nonlinear interaction
with the optical medium, this field gives rise to a nonlinear source polarization,
$\bm{\mathcal{P}}(\x,t) = \polNL(\x)\exp(i\Omega t) + c.c.$, which oscillates at the SH frequency,
$\Omega=2\omega_{0}$. This polarization is the source of the electromagnetic field generated at
the SH and is related to the excitation field via a third-order nonlinear susceptibility tensor,
$\bfchi$:
\begin{eqnarray}
    P_{\gamma}^{\mathrm{NL}} (\x) = \sum_{\alpha,\beta} \chi^{(2)}_{\gamma\alpha\beta} E_\alpha(\x) E_\beta(\x).
\label{eq:def_P_generic}
\end{eqnarray}
In this equation and what follows, Greek indices take the values of the Cartesian coordinates $x$,
$y$, and $z$.

Under the undepleted pump approximation, the initially nonlinear problem in the time-domain is
solved in three steps: \textit{i}) Calculate the field at the FF, $\omega_{0}$; \textit{ii})
evaluate the nonlinear polarization generated at the SH via Eq.~\eqref{eq:def_P_generic}; and
\textit{iii}) calculate the field at the SH. The first and last steps are performed using the
linear and nonlinear (extended) versions of the GSM, respectively.

\subsection{The GSM for inhomogeneous problems}
At both the FF and SH, the electromagnetic field is governed by the electromagnetic wave equation
with a physical source current, $\J^\text{ext}(\x)$,
\begin{eqnarray}
    \nabla \times \nabla \times \E - \omega^2\mu_0\varepsilon \E = i\omega\mu_0 \J^\text{ext},
\label{eq:def_waveequation}
\end{eqnarray}
where for convenience we have suppressed the spatial dependence of the variables and we introduced
a generic frequency, $\omega$, that takes the values $\omega = \omega_{0}$ ($\omega = \Omega$) at
the FF (SH). In Eq.~\eqref{eq:def_waveequation}, $\varepsilon(\x)$ is the spatial distribution of
the electric permittivity that defines the grating.

The GSM is a rigorous method for solving Eq.~\eqref{eq:def_waveequation}. It is based on the
decomposition of $\varepsilon(\x)$ defining the grating into a simple background structure with
permittivity, $\varepsilon_b(\x)$, and the difference structure, $\deps(\x) =
\varepsilon(\x)-\varepsilon_b(\x)$. The background has to be chosen such that the linear solution
operator $\Nb$ characterizing the background structure problem, namely the operator that
associates a source $\J$ with the corresponding solution of Eq.~\eqref{eq:def_waveequation} with
$\varepsilon = \varepsilon_b$, is known. This means that
\begin{eqnarray}
    \E = \Nb(\J)
\label{eq:def_Nb}
\end{eqnarray}
solves Eq.~\eqref{eq:def_waveequation} for $\J^\text{ext} = \J$ and $\varepsilon = \varepsilon_b$.

In order to find the fields for an arbitrary permittivity $\varepsilon(\x)$ one rewrites
Eq.~\eqref{eq:def_waveequation} as
\begin{eqnarray}
\nabla \times \nabla \times \E - \omega^2\mu_0 \varepsilon_b \E = i\omega\mu_0 \left[\J^\text{ext}
+ \J^\text{gen}(\E) \right],
\end{eqnarray}
where the term $\J^\text{gen}(\E)=-i\omega\deps\E$, which depends on the solution $\E$ itself, is called
generalized source. Using the solution operator this can be rewritten as
\begin{eqnarray}
    \E = \Nb\left(\J^{\mathrm{ext}} + \J^\text{gen}(\E)\right) =
    \Nb\left(\J^\text{tot}(\E)\right),
\label{eq:WE_impl_operator}
\end{eqnarray}
with $\J^\text{tot}(\E)=\J^{\mathrm{ext}} + \J^\text{gen}(\E)$. At the FF we choose the external
source term $\J^\text{ext}$ such that the external field, $\E^\text{ext}$, is an incident plane
wave, although other choices are possible. At the SH the external source term is given in terms of
the nonlinear polarization: $\J^\text{ext} = -i\Omega \polNL$.

\begin{figure}
\begin{center}
        \includegraphics[width=0.47\textwidth]{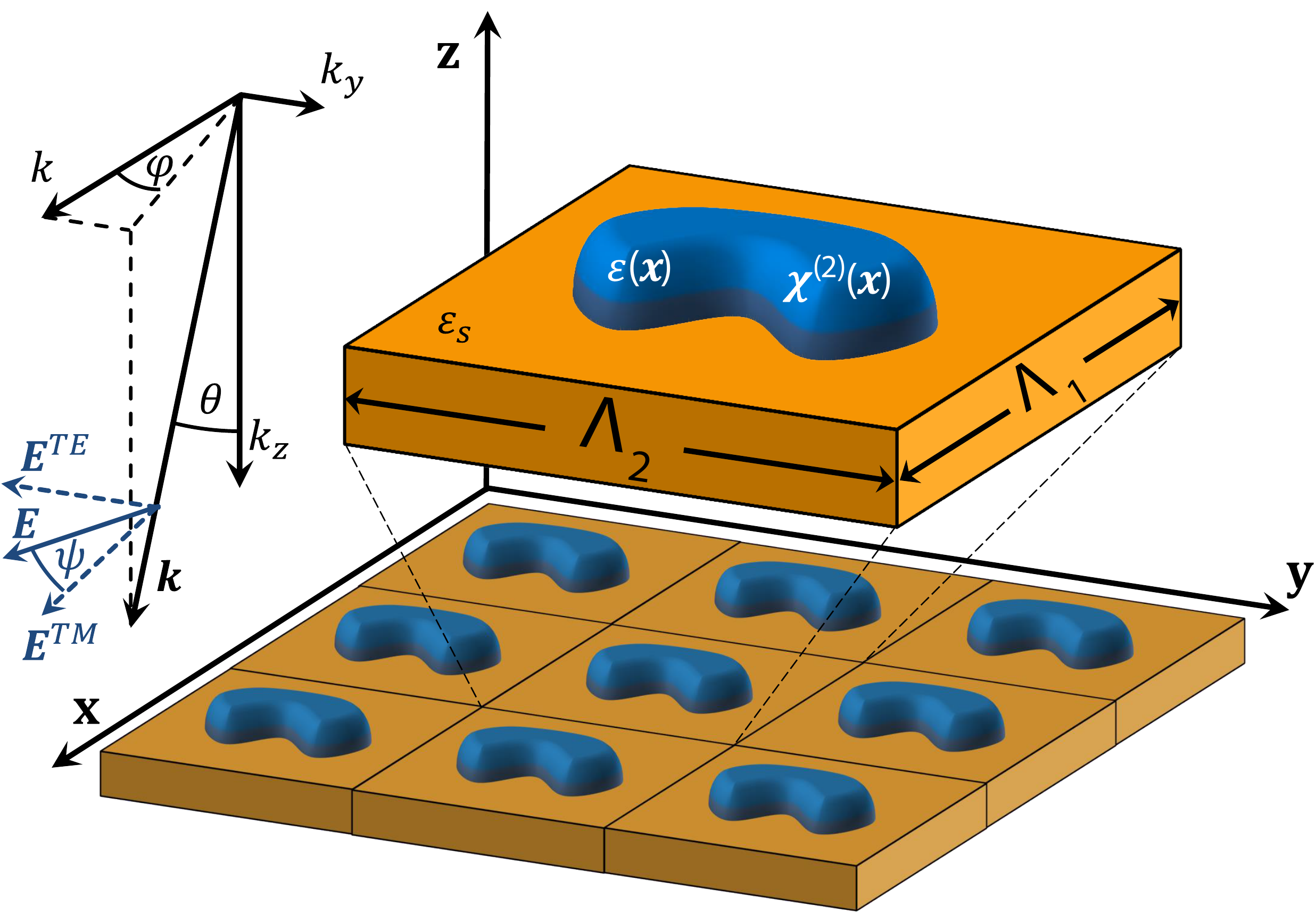}
\end{center}
\caption{\label{fig:setting} Setting for the GSM and closeup of the unit cell of a 2D corrugated
grating containing nonlinear material. The grating is described by $\varepsilon(\x)$ and
$\bfchi(\x)$ and is under plane wave incidence with wavevector $\k_0$.}
\end{figure}

So far, no particular assumptions about the structure under investigation have been made and thus
the formulation \eqref{eq:WE_impl_operator} is valid for any structure. We consider a 2D
corrugated grating, as depicted in \figref{fig:setting}. It consists of the grating region $0\leq
z\leq \height$, with grating vectors $\K_{1}$ and $\K_{2}$ lying in the transverse $(x,y)$ plane,
and a given permittivity distribution, $\varepsilon(\x)$, which is a periodic function of the $x$
and $y$ coordinates. This slab is sandwiched in-between the cover and substrate, which consist of
linear optical materials with permittivity $\varepsilon_c$ and $\varepsilon_s$, respectively. We
assume that only the grating region contains nonlinear material, i.e. $\bfchi=0$ if $z<0$ or
$z>\height$. Due to its periodicity, the permittivity of the structure can be decomposed in a
Fourier series,
\begin{eqnarray*}
    \varepsilon(\x) = \sum_{n=-\infty}^\infty \varepsilon_n(z) e^{i\left[ (n_1 K_{1x} + n_2 K_{2x}) x + (n_1K_{1y} + n_2 K_{2y})
    y\right]},
\end{eqnarray*}
where the sum over $n\equiv(n_1,n_2)$ is to be understood as a double infinite sum over the
integers $n_1$ and $n_2$, $n_{1,2}\in \mathbb{Z}$. Because $\varepsilon(\x)$ is a periodic
function in the transverse plane, one can use Bloch theorem to express the electric field as a
Fourier series of functions that are pseudo periodic on the transverse coordinates $x$ and $y$,
\begin{eqnarray}
  \E(\x) = \sum_{n=-\infty}^\infty \E_n(z) e^{i\left(k_{nx} x + k_{ny} y\right)},
\label{eq:def_E_everywhere}
\end{eqnarray}
where $k_{nx/y} = k_{0x/y} + n_1 K_{1x/y} + n_2 K_{2x/y}$ are the projections of the wavevector of
the $n^\text{th}$ order diffraction mode. The principal direction of the central diffraction order, which
is defined by $n_1=n_2=0$, is given by $\k_{0} =
k_c(\sin\theta\cos\phi,\sin\theta\sin\phi,\cos\theta)$, where $k_c =
\omega_{0}\sqrt{\varepsilon_c\mu_0}$ is the wavenumber in the cover region. For oblique incidence, this
leads to a phase-shift of the electric field $\E(\x)$ of $e^{i\left( k_{0x} \Lambda_1 + k_{0y}
\Lambda_2\right)}$ over a unit cell, where $\Lambda_1$ and $\Lambda_2$ are the periods along the
$x$ and $y$ axes, respectively.

% Since we do not include in our analysis nonlinear optical effects in the substrate and cover materials and
Since this analysis does not include nonlinear optical effects in the cover and substrate materials and since the nonlinear polarization $\polNL(\x)$ arises due to a periodically
distributed field \eqref{eq:def_E_everywhere} at the FF, both the external current $\J^\text{ext}
= -i\Omega \polNL$ at the SH and the generalized source term can be expressed as Fourier series,
\begin{eqnarray}
  \J^\text{tot} &=& -i\Omega \polNL -i\Omega(\varepsilon-\varepsilon_b)\E \nonumber  \\
    &=& \sum_{n=-\infty}^\infty \left[\j^\text{NL}_n(z)+\j^{\mathrm{gen}}_n(z)\right] e^{i(k_{nx}x+k_{ny}y)}.
\label{eq:j_tot}
\end{eqnarray}
The coefficients $\j^\text{gen}_n(z)$ depend on all field terms $\E_m(z)$
\begin{eqnarray*}
  \j^\text{gen}_n(z) &=& -i\omega \left\{\left[\varepsilon(z)-\varepsilon_b(z)\right]\E(z)\right\}_n \nonumber \\
                                         &=& -i\omega \sum_{m=-\infty}^\infty\left[\varepsilon(z)-\varepsilon_b(z)\right]_{n-m}\E_m(z).
\end{eqnarray*}

The solution operator $\Nb$ for a periodic source current, $\j_n(z)e^{i(k_{nx}x+k_{ny}y)}$, in
homogeneous space with constant permittivity $\varepsilon_b$ reads \cite{st12jqsrt,s87josab}:
\begin{eqnarray}
    \E_n(z) = \frac{i}{\omega}\int \Gf_n(z,z') \j_n(z')dz', \label{eq:Green}
\end{eqnarray}
with the tensor Green's function
\begin{align}
    \Gf_n(z,z') =&\left({\te^+_n}{\te^+_n}^T + \tm^+_n{\tm^+_n}^T\right)\heavi(z\minus z')e^{i\kappa^b_{nz} (z\minus z')} \nonumber \\
                      + &\left({\te^-_n}{\te^-_n}^T + \tm^-_n{\tm^-_n}^T\right)\heavi(z'\minus z)e^{i\kappa^b_{nz} (z'\minus z)} \nonumber \\
                      -  &\varepsilon_b^{-1} \ez\ez^T \delta(z-z').
\label{eq:GreenS}
\end{align}
Here, $\te_n^\pm$ and $\tm_n^\pm$ are TE and TM component vectors, respectively, of the
$n^\text{th}$ Fourier term of the electric field $\E_n$, $H(\cdot)$ denotes the Heaviside step
function, and $\mathbf{a}^{T}$ means the transpose of a vector $\mathbf{a}$. \uline{The unit vector in $z$-direction is denoted by $\ez$}. Also, the dispersion
relation
$\kappa^{b}_{nz}=\left[\mu_0\varepsilon_b\omega^{2}-(k_{nx})^{2}-(k_{ny})^{2}\right]^{1/2}$ holds.
One can see that the tensor $\Gf_n$ gives rise to upwards and downwards propagating plane waves
and a stationary term. In order to express the solution in terms of a superposition of plane
waves, we eliminate the contribution of the $\delta$-term in Eq.~\eqref{eq:GreenS} by defining the
modified electric field
\begin{eqnarray}
    \tilde E_\alpha(z) = E_\alpha(z) - \frac{\delta_{\alpha,z}}{i\omega\varepsilon_b} J_\alpha^\text{tot}(z). \label{eg:defTildeE}
\end{eqnarray}
The Fourier components of this modified field, $\tilde \E_n(z) = \textsf{Q}_n\amp_n(z)$, can be
expressed via their TE/TM amplitudes, $\amp_n(z)=\left(a_{en}^+(z),a_{en}^-(z),a_{hn}^+(z),
a_{hn}^-(z)\right)$, and the matrix $\textsf{Q}_n=\left(\te_n^+,\te_n^-,\tm_n^+,\tm_n^-\right)$.
Additionally, Eq.~\eqref{eg:defTildeE} leads to a useful property of its $z$-component:
\begin{eqnarray}
    \tilde E_z(z) = \frac{D_z}{\varepsilon_b}.
\end{eqnarray}

The total source can be rewritten as
\begin{eqnarray}
    \J^\text{tot} &=& \J^\text{ext} + \J^\text{gen}(\E) = \J^\text{ext} - i\omega (\varepsilon \E - \varepsilon_b \E ) \nonumber  \\
        &=&-i\omega (\P^\text{NL} + \varepsilon \E - \varepsilon_b \E) = -i\omega \left(\D - \D_b
        \right),
        \label{eq:Jtot}
\end{eqnarray}
where $\D_b=\varepsilon_b \E$ and $\P^\text{NL}=0$ if $\omega=\omega_{0}$.
So far, an \emph{infinite} number of terms in the Fourier series was considered.
% So far, we assumed an \emph{infinite} number of terms in the Fourier series.
However, it is well known
(\cite{lm96josaa,l96josaa,l97josaa}) that the factorization of products of periodic functions with a
\emph{finite} number of terms has to be performed with care, namely the correct approach is
dictated by the continuity properties of the factors in the product. To be more specific, products
of functions with non-simultaneous discontinuities (e.g., the tangential component of the electric
displacement field $D_\parallel = \varepsilon E_\parallel + P^\text{NL}_\parallel$) can be
factorized using Laurent's product rule, whereas continuous products of functions with
simultaneous discontinuities (the normal component $D_\perp=\varepsilon E_\perp +
P^\text{NL}_\perp$ is continuous at interfaces, whereas $\varepsilon$ and $E_\perp$ can be
discontinuous) can be factorized using the inverse rule. In the context of numerical methods for
periodic structures in the Fourier space, such as the FMM and GSM, the violation of the correct
factorization rules causes slow convergence with respect to the number of retained diffraction
orders. In contrast to previous work \cite{wgp14spie}, where the correct factorization was only
used in the linear part of the calculations, we now show how the correct factorization should be
implemented in the nonlinear part of the calculations.

\begin{figure}
    \centering
    \includegraphics[width=8cm]{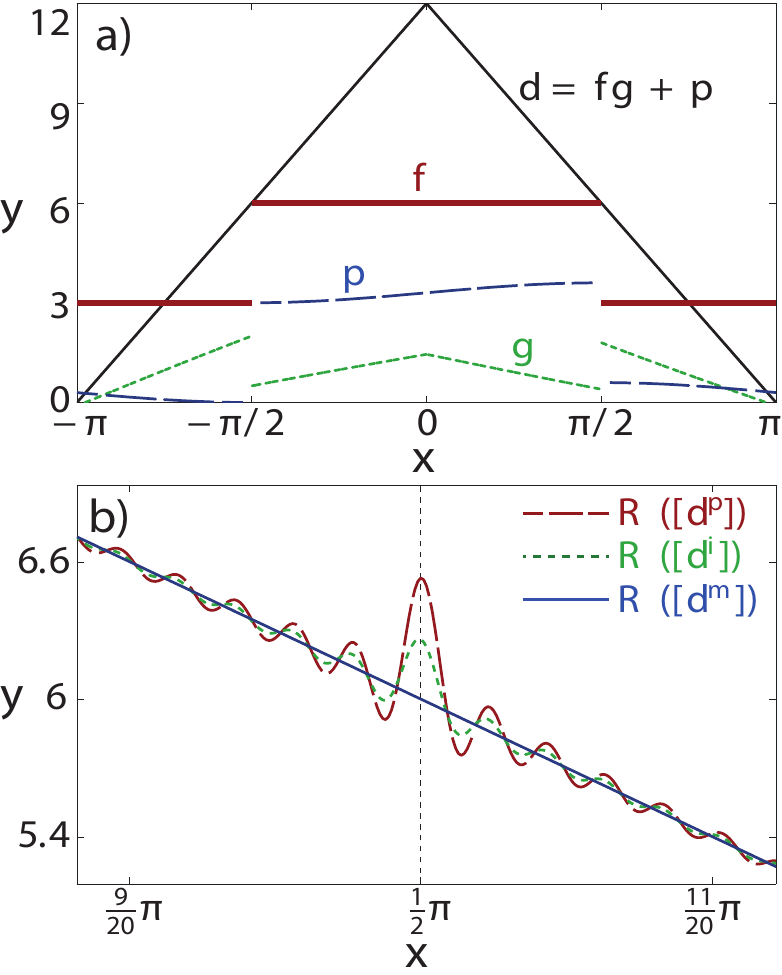}
\caption{a) Discontinuous functions $f$ (thick brown), $g$ (dashed green), and $p$ (broken blue lines)
and continuous function $d$ (solid black), for $a=6$, $b=2$, $c=3$. b) Reconstruction of Fourier
series with $200$ components around discontinuity obtained by the product rule (broken brown lines), inverse
rule (dashed green), and modified inverse rule (solid blue).\label{fig:ExampleLi}}
\end{figure}

We now present the derivation of this modified rule and the impact of its correct application on
the accuracy of the calculations is investigated in Section~\ref{sec:Benchmark}. Thus, to
illustrate why a modified factorization rule for the inhomogeneous wave equation is necessary, we
consider the example given in \cite{l96josaa}, adapted to our case:
\begin{eqnarray*}
    f(x) &=& \left\{\begin{array}{ll} a, & |x|<\pi/2,\\
        a/2, & |x|\geq\pi/2, \end{array}\right.\\
    p(x) &=& 0.1c[1+\sin(x)] + \left\{\begin{array}{ll} c, & |x|<\pi/2,\\
        0, & |x|\geq\pi/2, \end{array}\right.\\
    g(x) &=& - \frac{p(x)}{f(x)} + \left\{\begin{array}{ll} b(1-|x|/\pi), & |x|<\pi/2,\\
        2b(1-|x|/\pi), & |x|\geq\pi/2, \end{array}\right. \\
    d(x) &=& f(x)g(x) + p(x).
\end{eqnarray*}
Here, $a,b>0$ and $c\geq 0$ are arbitrary constants. One can easily see that the function $d$ is
continuous, whereas $f$ and $g$ have concurrent discontinuities at $|x|=\pi/2$. For $c\neq 0$, $p$
is also discontinuous at $|x|=\pi/2$. Otherwise, Li's example \cite{l96josaa} is obtained.
Figure~\ref{fig:ExampleLi}(a) shows the functions for $a=6$, $b=2$, and $c=3$.

In order to obtain the vector, $\fvec{d}$, of Fourier coefficients of $d$, three different
factorizations are compared, the results being depicted in \figref{fig:ExampleLi}(b). The product
rule yields $\fvec{d^p} = \fmatrix{f}\fvec{g} + \fvec{p}$, where $\fmatrix{f}$ is the Toeplitz
matrix of the Fourier coefficients of $f$. The reconstruction, $\rec{\fvec{d^p}}$, of the function
$d$ shows strong oscillations at $|x|=\pi/2$ around the expectedly continuous function $d$ itself.
The reason for this behavior is that the product rule propagates the overshoot from
$\rec{\fvec{f}}$ and $\rec{\fvec{g}}$ to the reconstruction of the product
$\rec{\fmatrix{f}\fvec{g}}$. In addition, the oscillatory behavior of $\rec{\fvec{p}}$ at
$|x|=\pi/2$ adds to the variations of the reconstructed product, $\rec{\fvec{d^p}}$.

A direct application of the inverse rule for the product yields $\fvec{d^i} =
\fmatrix{1/f}^{-1}\fvec{g} + \fvec{p}$. The oscillations of the reconstruction $\rec{\fvec{d^i}}$
are now less pronounced but still consist of additive contributions from both terms
$\rec{\fmatrix{1/f}^{-1}\fvec{g}}$ and $\rec{\fvec{p}}$.

We propose a modification of the product rule: we rewrite the continuous function $d(x)=f(x)g(x) +
p(x)=f(x)\big[g(x) + p(x)/f(x)\big]$ as a product of discontinuous functions \textit{in the
space-domain}. The two factors have complementary discontinuities, whence the inverse rule applied
to this product yields the correct factorization, $\fvec{d^m} = \fmatrix{1/f}^{-1}\left(\fvec{g} +
\fvec{p/f}\right)$. This is illustrated by the reconstruction $\rec{\fvec{d^m}}$, which shows no
spurious oscillations at the points of discontinuity.

Armed with this correct factorization rule, let us now consider Eq.~\eqref{eq:Jtot}. Thus, we find
that in the context of the nonlinear GSM, $d$, $f$, $g$, and $p$ correspond to the continuous
normal component of the displacement field, $D_\perp$, the permittivity, $\varepsilon$, $E_\perp$,
and $P^\text{NL}_\perp$, respectively. The correct factorization of tangential and parallel
components of $\D$ therefore reads:
\begin{subequations}
\begin{align}
    \fvec{D_\parallel} &= \fmatrix{\varepsilon} \fvec{E_\parallel} + \fvec{P^\text{NL}_\parallel}, \\
    \fvec{D_\perp} &= \fmatrix{1/\varepsilon}^{-1} \left( \fvec{E_\perp} + \fvec{P^\text{NL}_\perp /\varepsilon
    }\right).
\end{align}
\end{subequations}
We specify again that $\fvec{V}$ denotes the vector of $N_o=(2N_1+1)(2N_2+1)$ Fourier coefficients
of $V$. The truncation of Fourier terms with $n_1=-N_1,\ldots, N_1$ and $n_2=-N_2,\ldots, N_2$,
with $N_{1}$ and $N_{2}$ integers, corresponds to a rectangular truncation pattern in the Fourier
space. Problem specific, non-rectangular patterns could also be used and may lead to faster
convergence in certain cases (see \cite{l97josaa} for a study of different truncation patterns in
the FMM). Accordingly, the Fourier coefficients of the total source current, $\j_n =
\j_n^\text{NL} + \j_n^\text{gen}$, defined by Eq.~\eqref{eq:j_tot} are given by
\begin{eqnarray*}
    \fvec{j_\parallel} &=& -i\omega \varepsilon_b \left\{ \left(\fmatrix{\frac{\varepsilon}{\varepsilon_b}} - \textsf{I}\right) \fvec{E_\parallel} + \varepsilon_b^{-1} \fvec{P^\text{NL}_\parallel}\right\},\\
    \fvec{j_\perp} &=& -i\omega \varepsilon_b \left\{ \left(\fmatrix{\frac{\varepsilon_b}{\varepsilon}}^{-1}\!\!\!\! - \textsf{I}\right) \fvec{E_\perp} + \fmatrix{\frac{\varepsilon_b}{\varepsilon}}^{-1}\fvec{P^\text{NL}_\perp/\varepsilon}\right\},
\end{eqnarray*}
\uline{where $\textsf{I}$  denotes the identity matrix of respective size.}

It remains to reformulate these equations, which are given in terms of the
$(\perp,\parallel)$-components of $\E$, in terms of the Cartesian components of $\tilde \E$, the
latter ones being the actual variables used in the GSM formalism. To this end, a local
normal/tangential vector field is defined so that for any vectorial quantity $\bf v$ we can write:
\begin{eqnarray*}
    \left(\begin{array}{c} v_{x} \\ v_{y} \\ v_{z} \end{array}\right) = \textsf{B}\left(\begin{array}{c} v_{n} \\ v_{\psi} \\ v_{\phi} \end{array}\right),
\end{eqnarray*}
where the orthogonal transformation matrix
\begin{eqnarray*}
    \textsf{B} = \left(\hat{\mathbf{n}},\hat{\mathbf{\psi}}, \hat{\mathbf{\phi}} \right) =
        \left( \begin{array}{ccc} \cos\phi\sin\psi & \cos\phi\cos\psi & -\sin\phi \\
                                                             \sin\phi\sin\psi & \sin\phi\cos\psi &  \cos\phi \\
                                                             \cos\psi & -\sin\psi &  0
                                                            \end{array}\right)
\end{eqnarray*}
is a concatenation of the unit vectors in normal, and both tangential directions $\hat {\bf n}$,
$\hat{\mathbf{\psi}}$, and $\hat{\mathbf{\phi}}$, respectively. Using this transformation we get
(analogously to Appendix A in \cite{st12jqsrt}):
\begin{eqnarray}
    \fvec{j_\alpha} = -i \omega \varepsilon_b \sum_\beta \left(\Delta \delta_{\alpha\beta} - \textsf{D}\Gamma_{\alpha\beta}\right) \fvec{E_\beta}
       -i \omega \varepsilon_b \fvec{\bar p_\alpha^{p}}, \ \ \ \
        \label{eq:JasE}
\end{eqnarray}
where
\begin{eqnarray*}
    \left(\begin{array}{c} \fvec{\bar p^{p}_x} \\ \fvec{\bar p^{p}_y}\\ \fvec{\bar p^{p}_z}\end{array}\right) =
    \textsf{B}\ \mathrm{diag}\left(\fmatrix{\frac{\varepsilon_b}{\varepsilon}}^{-1},\varepsilon_b^{-1},\varepsilon_b^{-1}\right)
    \left(\begin{array}{c} \fvec{p^{p}_n/\varepsilon} \\    \fvec{p^{p}_\psi}\\ \fvec{p^{p}_\phi}\end{array}\right).
\end{eqnarray*}
The matrices $\Delta$ and \textsf{D} are defined as
\begin{eqnarray*}
    \Delta = \fmatrix{\frac{\varepsilon}{\varepsilon_b}} - \textsf{I} \quad \text{and} \quad
    \textsf{D} = \fmatrix{\frac{\varepsilon}{\varepsilon_b}} - \fmatrix{\frac{\varepsilon_b}{\varepsilon}}^{-1}
\end{eqnarray*}
and $\Gamma_{\alpha\beta}$ are the Fourier images of the entries of the matrix
{\small
\begin{eqnarray*}
    \!\Gamma(z)\! = \!\!\left(\!\!
    \begin{array}{ccc}
    \cos^2\!\phi\sin^2\!\psi & \sin\phi\cos\phi\sin^2\!\psi & \cos\phi\sin\psi\cos\psi\\
    \sin\phi\cos\phi\sin^2\!\psi & \sin^2\!\phi\sin^2\!\psi & \sin\phi\sin\psi\cos\psi\\
    \cos\phi\sin\psi\cos\psi   & \sin\phi\sin\psi\cos\psi & \cos^2\!\psi
    \end{array}\!\!\right)\!\!.
\end{eqnarray*}
}
\uline{In agreement with the modified factorization rule, the Fourier series coefficients $\fvec{p^p_n/\varepsilon}$ are obtained by numerical integration of the space-domain quotient of the functions $p^p_n(x)$ and $\varepsilon(x)$.}

In order to eliminate $\E$ in favor of $\tilde \E$ we insert
\begin{equation*}
    \frac{\fvec{j_z}}{i\omega \varepsilon_b} =  \textsf{D}\left(\Gamma_{zx}\fvec{E_x} + \Gamma_{zy}\fvec{E_y}\right) + \left(\textsf{D}\Gamma_{zz}-\Delta\right)\fvec{E_z}-\fvec{\bar
    p^{p}_z}
\end{equation*}
into the definition \eqref{eg:defTildeE} of $\tilde \E$ to obtain
\begin{equation*}
    \tilde{\fvec{E_z}} = -\textsf{D}\left(\Gamma_{zx}\fvec{E_x} + \Gamma_{zy}\fvec{E_y}\right) + \textsf{C} E_z + \fvec{\bar p^{p}_z},
\end{equation*}
where $\textsf{C}=\Delta-\textsf{D}\Gamma_{zz}$. This yields $\fvec{\E}$ resolved in terms of
$\tilde{\fvec{\E}}$, which in matrix form is expressed as
\begin{eqnarray}
    \left(\begin{array}{c} \fvec{E_x} \\ \fvec{E_y}\\ \fvec{E_z}\end{array}\right)
     &=&  \left(\begin{array}{ccc} \textsf{I} & 0 & 0 \\ 0 & \textsf{I} & 0 \\ \textsf{C}^{-1}\textsf{D}\Gamma_{zx} & \textsf{C}^{-1}\textsf{D}\Gamma_{zy} & \textsf{C}^{-1}\end{array}\right)  { \left(\begin{array}{c} \tilde{\fvec{E_x}} \\ \tilde{\fvec{E_y}}\\ \tilde{\fvec{E_z}}\end{array}\right) } \nonumber \\
    &-&\left(\begin{array}{c} 0 \\ 0\\ \textsf{C}^{-1} \fvec{\bar p^{p}_z}\end{array}\right).
\end{eqnarray}
Plugging this formula back into Eq.~\eqref{eq:JasE} provides the total source current, which
consists of the generalized source dependent on the unknown field $\tilde \E$ and a physical
source whose origin is the known polarization:
\begin{eqnarray}
    \frac{\fvec{j_{\alpha}}}{i\omega \varepsilon_b}= \sum_\beta \textsf{W}_{\alpha\beta} \tilde{\fvec{E_{\beta}}} + \fvec{\tilde
    p_{\alpha}^{p}}, \label{eq:jFinal}
\end{eqnarray}
where the modified polarization reads
\begin{eqnarray}
    \fvec{\tilde p_\alpha^{p}} &=& - \fvec{\bar p^{p}_\alpha} + \left(\Delta \delta_{\alpha z} - \textsf{D}\Gamma_{\alpha z}\right) \textsf{C}^{-1} \fvec{\bar p^{p}_z}
    \label{eq:tildePFinal}
\end{eqnarray}
and the matrix $\textsf{W}(z)$ is defined as
\begin{equation*}
    \textsf{W} =
    \left( { \begin{array}{ccc}
    \textsf{N}_{xx}+\textsf{D}\Gamma_{xx}-\Delta &       \textsf{N}_{xy}+\textsf{D}\Gamma_{xy} & \textsf{D}\Gamma_{xz}\textsf{C}^{-1}\\
         \textsf{N}_{yx}+\textsf{D}\Gamma_{xx} & \textsf{N}_{xy}+\textsf{D}\Gamma_{yy}-\Delta & \textsf{D}\Gamma_{yz}\textsf{C}^{-1}\\
                 -\textsf{C}^{-1}\textsf{D}\Gamma_{zx}  &  -\textsf{C}^{-1}\textsf{D}\Gamma_{zy}        & \textsf{C}^{-1}-\textsf{I}
    \end{array}} \right). \label{eq:defW}
\end{equation*}
Here, $\textsf{N}_{\alpha\beta} = \textsf{D}\Gamma_{\alpha
z}\textsf{C}^{-1}\textsf{D}\Gamma_{z\beta}$ combines the geometrical normal field information
contained in $\Gamma$ with the physical structure given by $\varepsilon$.

Combining the main results, namely Eq.~\eqref{eq:jFinal} that expresses the source in terms of
$\tilde \E$ and the Green's function \eqref{eq:Green}, yields a system of Fredholm integral equations of
the second kind for the unknown amplitudes,
\begin{eqnarray}
    {\bf a}_n(z) &=& {\bf a}^0_n(z) + \int_{z'=0}^\height \textsf{R}_n(z,z') \sum_{\alpha}\bar{\textsf{Q}}_{:\alpha,n} \nonumber \\
        &\times& \sum_{m=-N_o}^{N_o} \sum_\beta \textsf{W}_{\alpha\beta,nm}(z')\textsf{Q}_{\beta:,m} {\bf a}_{m}(z') dz', \label{eq:def_sol_n}
\end{eqnarray}
for $n=1,\ldots,N_o$, where the sum over $m\equiv(m_1,m_2)$ is to be understood as a double finite sum over the integers $m_1=-N_1,\ldots, N_1$ and $m_2=-N_2,\ldots, N_2$.
\uline{$\bar{\textsf{Q}}_{:\alpha,n}$ and $\textsf{Q}_{\beta:,m}$ denote the $\alpha$-column and $\beta$-row of $\bar{\textsf{Q}}_{n}$ and $\textsf{Q}_{m}$, respectively.}
The matrix
$\textsf{R}_n(z,z')\in\mathbb{C}^{4\times 4}$ incorporates the propagation of TE- and
TM-polarized, upward and downward plane wave amplitudes of the $n^\text{th}$ Fourier component
inside the slab-background structure and their reflections at the top and bottom interfaces [see
Eqs.~(61) and (62) in \cite{st12jqsrt} for the exact formula]. The known values of ${\bf
a}_n^0(z)$ are either the amplitudes of the incident plane wave electric field in the background
structure at the FF or the amplitudes of the external source field $\Nb(\mathbf{J}^\text{ext})$ at
the SH; the latter are given by
\begin{eqnarray}
    {\bf a}^0_n(z) = \int_{z'=0}^\height \textsf{R}_n(z,z') \bar{\textsf{Q}}_n\tilde\p^p_n(z') dz'.
    \label{eq:def_Nb0_n}
\end{eqnarray}

One can easily see that the modified factorization rule does not affect the generalized source but
only changes the external polarization. This means, that all algorithmic enhancements, that make
the linear GSM an efficient numerical method, can be applied here as well.

Using the midpoint rule, the system \eqref{eq:def_sol_n} of integral equations is discretized at
$N_l$ equidistant points, $z_p = (p-0.5)\height/N_l$, $p=1,\ldots,N_{l}$. This results in a system
of linear equations,
\begin{eqnarray}
    {\bf a}_{np} &=& {\bf a}^0_{np} + \frac{\height}{N_l}\sum_{q=1}^{N_l} \textsf{R}_n(z_p,z_q) \sum_{\alpha} \bar{\textsf{Q}}_{:\alpha,n} \nonumber \\
        &\times&  \sum_{m=-N_{o}}^{N_{o}} \sum_\beta \textsf{W}_{\alpha\beta,nm}(z_q)\textsf{Q}_{\beta:,m} {\bf a}_{mq}, \label{eq:def_sol_np}
\end{eqnarray}
for $n=1,\ldots,N_0, p=1,\ldots,N_l$. The total number of unknowns in Eq.~\eqref{eq:def_sol_np} is $4N_{o} N_l$ and usually becomes prohibitively large for direct solvers to be used in practical
situations of interest.

In order to take full advantage of an iterative solver, the system \eqref{eq:def_sol_np} is
reformulated to (see also \cite{st12jqsrt}):
\begin{equation}
    {\bf a} = \left(\textsf{I} + \textsf{R}\bar{\textsf{Q}}\textsf{U} \textsf{A}^{-1}\textsf{Q}\right){\bf a}^0, \label{eq:sol_iter}
\end{equation}
where ${\bf a}$ and ${\bf a}^0$ are the vectors containing the discrete unknowns, ${\bf a}_{np} =
{\bf a}_n(z_p)$, and the known amplitudes, ${\bf a}^0_{np} = {\bf a}^0_n(z_p)$, respectively,
$\textsf{A}=\textsf{M}-\textsf{QR}\bar{\textsf{Q}}\textsf{U}$, and $\bar{\textsf{Q}}$ and
$\textsf{Q}$ are block-diagonal matrices with $\bar{\textsf{Q}}_n$ and $\textsf{Q}_n$ on their
block-diagonal, respectively. The matrix $\textsf{R}\in\mathbb{C}^{4N_oN_l \times 4N_oN_l}$
corresponds to $\textsf{R}_n(z_p,z_q)$: Its $\big(4(n-1)N_l+4(p-1)+[1,..,4], \
4(n-1)N_l+4(q-1)+[1,..,4]\big)$-entries are given by $\textsf{R}_n(z_p,z_q)$. Hence, $\textsf{R}$
is diagonal with respect to the Fourier component $n$ and has Toeplitz-structure with respect to
the layer indices $p$ and $q$. The matrices
\begin{small}
\begin{align}
    &\textsf{M}= \left(\!\!
    \begin{array}{ccc}
    \fmatrix{\frac{\varepsilon_b}{\varepsilon}}\fmatrix{\frac{\varepsilon}{\varepsilon_b}} & 0 & 0 \\
        0 & \fmatrix{\frac{\varepsilon_b}{\varepsilon}}\fmatrix{\frac{\varepsilon}{\varepsilon_b}} & 0 \\
        0 & 0 & \fmatrix{\frac{\varepsilon}{\varepsilon_b}}\fmatrix{\frac{\varepsilon_b}{\varepsilon}}
    \end{array}\!\!\right)\small \sin^2 \psi + \cos^2 \psi, \nonumber\\
&\textsf{F}=\textsf{I}-\fmatrix{\frac{\varepsilon_b}{\varepsilon}}\fmatrix{\frac{\varepsilon}{\varepsilon_b}},~~~
\textsf{G}=\textsf{I}-\fmatrix{\frac{\varepsilon}{\varepsilon_b}}\fmatrix{\frac{\varepsilon_b}{\varepsilon}},\nonumber\\
&\textsf{U} = \left(\!\!
    \begin{array}{ccc}
    \Delta\textsf{M}_{xx} + \textsf{G}\fmatrix{\frac{\varepsilon}{\varepsilon_b}}\Gamma_{xx} \!\!\! &
                                                        \textsf{G}\fmatrix{\frac{\varepsilon}{\varepsilon_b}}\Gamma_{xy} &
                                                        \textsf{G}\Gamma_{xz} \\
                                                        \textsf{G}\fmatrix{\frac{\varepsilon}{\varepsilon_b}}\Gamma_{yx} &
       \!\!\! \Delta\textsf{M}_{yy} + \textsf{G}\fmatrix{\frac{\varepsilon}{\varepsilon_b}}\Gamma_{yy} &
                                                        \textsf{G}\Gamma_{yz} \\
                                                        \textsf{F}\Gamma_{zx} &
                              \textsf{F}\Gamma_{zy} &
                                                        \!\!\textsf{M}_{zz} - \fmatrix{\frac{\varepsilon_b}{\varepsilon}}
    \end{array}\!\!\right),\nonumber
\end{align}
\end{small}

are defined such that $\textsf{W}=\textsf{UM}^{-1}$ but they do not contain inverted submatrices
themselves. Additionally, $\textsf{U}$ and $\textsf{M}$ are block-Toeplitz-Toeplitz-matrices with
respect to the index pair $n=(n_1,n_2)$. Due to the Toeplitz-property of its submatrices,
multiplication of a vector ${\bf b}$ of size $3N_oN_l$ with the $3N_oN_l\times 3N_oN_l$ system
matrix $\textsf{A}$ can be performed in $\mathcal O(N_oN_l \log(N_o)\log(N_l))$ \cite{blahut85aw}
operations instead of $\mathcal O(N_o^2 N_l^2)$ for the standard matrix-vector multiplication.
Therefore, the use of an iterative solver based on matrix-vector multiplications like \nolinebreak
GMRES is highly beneficial. \uline{Although the reformulation Eq.~\eqref{eq:sol_iter} removes all
inverted submatrices from the system matrix $\textsf{A}$ itself, the evaluation of the modified
source polarization $\fvec{\tilde \p^p}$ in Eq.~\eqref{eq:tildePFinal} still requires a one-time
inversion of matrix $\textsf{C}$. In practice, the computational cost of this nonrecurring
inversion is, however, negligible compared to the overall runtime of the algorithm, which is
dominated by the inversion-free iterative solution process.}

%%%%%%%%%%%%%%%%%%%%%%%%%%%%%%%%%%%%%%%%%%%%%%%%%%%%%%%%%%%%%
\section{Numerical analysis and convergence studies}
\label{sec:Benchmark} In this section we examine the convergence and runtime properties of the
linear and nonlinear parts of the GSM using cylindrical gratings as test examples 2D made either
of dielectric or metallic materials. Importantly, we illustrate the impact of the application of
the correct factorization rule as compared to the old factorization.

The unit cell of the grating under consideration is assumed to be rectangular and, for simplicity,
the two periods are set to be equal, $\Lambda_1=\Lambda_2=\Lambda=$~\SI{2}{\micro\meter}. It
contains a cylindrical disk of height, $h=$~\SI{200}{\nano\meter}, and radius,
$r=0.25\Lambda=$~\SI{500}{\nano\meter}. The disk and substrate have refractive index, $n_{s}=2$,
whereas the cover region is vacuum with $n_{c}=1$. We assume an isotropic nonlinear susceptibility
tensor, $\chi^{(2)}_{jkl} = \bar{\chi}\delta_{jk}\delta_{kl}$, with
$\bar{\chi}=$~\SI{e-8}{\meter\per\volt}. The incident plane wave at the fundamental wavelength of
$\lambda_{\mathrm{FF}}=$~\SI{1.5}{\micro\meter}, is specified by the angles $\theta=$~\ang{30} and
$\phi=$~\ang{70}, and a $\psi=$~\ang{45} polarization.

The optical response of the device at the FF and SH is computed by the linear and nonlinear GSM, respectively.
We perform simulations for an increasing number of diffraction orders $N_{o}=(2N_1+1)(2N_2+1)$ for
$N_1=N_2\in \{3,5,...,23\}$ and increasing number of layers, $N_l\in\{12,25,50,...,400\}$. In
order to assess the efficiency of the algorithm, we compute the following quantities: the
reflection and transmission coefficients, $R^{FF/SH}$ and $T^{FF/SH}$, respectively, at the FF/SH,
the number $\Niter^{FF}$ and $\Niter^{SH}$ of GMRES iterations necessary to solve
Eq.~\eqref{eq:sol_iter} at the FF and SH, respectively, and the GSM runtime needed to complete the
linear ($t^{FF}$) and nonlinear ($t^{SH}$) part of the computations.

For the validation of the linear calculations results, they are compared to data obtained by using an
efficient normal vector field formulation \cite{srk07josaa} of the FMM, which is also commercially available \cite{omnircwaPD}. % which is particularly well suited for rounded grating structures.
For the presented comparison, both methods were implemented in MATLAB and all simulations
were performed on a single \SI{2.5}{\giga \hertz} processor to allow a fair comparison of simulation times.

%For the validation of the linear calculations, the results are compared to data obtained using an
%efficient normal vector field formulation \cite{srk07josaa} of the FMM, which is particularly well
%suited for rounded grating structures.
%Both methods were implemented in MATLAB and all simulations
%were performed on a single \SI{2.5}{\giga \hertz} processor for a fair comparison of simulation times.

\begin{figure}
    \centering
    \includegraphics[width=8cm]{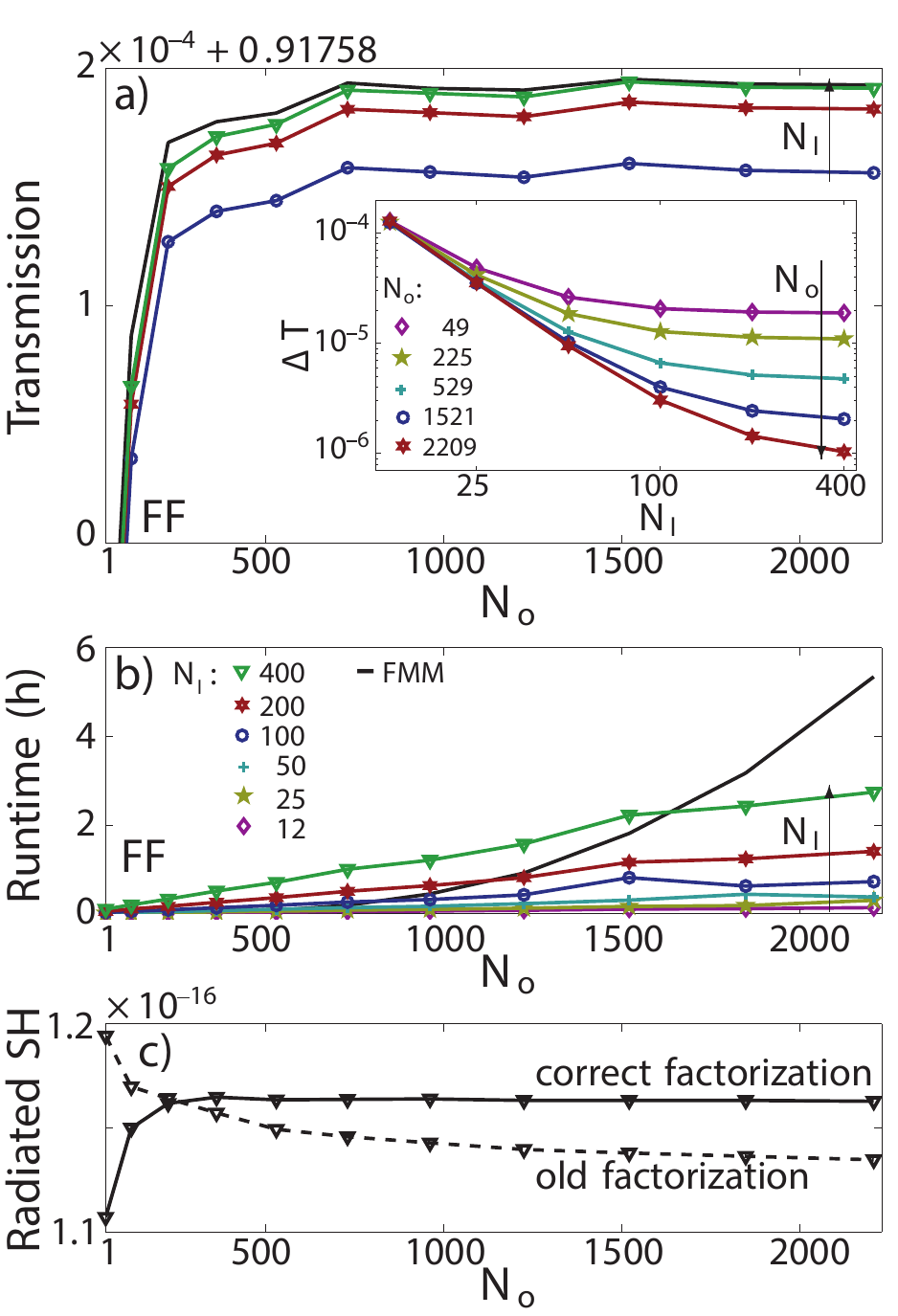}
\caption{\label{fig:convergence} a) Transmission efficiency obtained using GSM for increasing
$N_l$ [colors, legend from b)] and FMM (black). Inset: discretization error of GSM vs. $N_o$. b)
Runtime of GSM for increasing $N_l$ (colors) and of FMM (black). c) Convergence behavior of
radiated SH in the direction of the incoming plane wave using the two different factorization
rules.}
\end{figure}

The results of this analysis are summarized in \figref{fig:convergence}. The dependence of the
transmission coefficient at the FF, $T^{FF}$, on the number of diffraction orders, computed by
using the GSM for several values of the numbers of layers $N_l$ in which the grating is divided, is
presented in \figref{fig:convergence}(a). For comparison, we present the same dependence
determined by using the FMM. Both methods show a similar convergence pattern with respect to $N_o$
and the computed transmission coefficient rapidly approaches the asymptotic value of $T^{FF} =
0.917772$. For the investigated lossless device the GSM conserves the incoming optical power up to
a relative error of $10^{\minus 5}$ for low $N_o<1000$ and the power conservation improves to
$10^{\minus 6}$ as $N_o$ further increases.

The inset in \figref{fig:convergence}(a) illustrates the error in the transmission due to the
discretization of the integral equation, defined as $\Delta T = T^{FF} - T^{FF}_{FMM}$, calculated
for several values of orders, $N_o\in\{49,...,2209\}$, and for increasing $N_l$. Given
sufficiently many orders, e.g. $N_o=2209$, quadratic decrease of $\Delta T$ is achieved, which is
to be expected since using the midpoint rule in the discretization of the integral equation
implies quadratic convergence. For smaller $N_o$, e.g. $N_o=225$, the convergence plateaus at a
small number of layers. This is because the reformulated linear system \eqref{eq:sol_iter}, which
enables the use of the accelerated matrix-vector multiplication, is only $N_o$-asymptotically
equivalent to the original system \eqref{eq:def_sol_np}.

Figure~\ref{fig:convergence}(b) depicts the runtime $t^{FF}$ necessary to compute the solution of
the problem at the FF for increasing $N_o$ and for different $N_l$. The linear GSM exhibits nearly
optimal runtime as $t^{FF}$ increases log-linearly with both $N_o$ and $N_l$. This is because
$t^{FF}$ is chiefly determined by the computational work needed to iteratively find the solution
of Eq.~\eqref{eq:sol_iter} using GMRES, which in turn depends on the number of iterations
$\Niter^{FF}$ and, if $\Niter^{FF} \ll N_oN_l$, on the time required for the system-matrix vector
multiplication; this latter quantity is of order $\mathcal O(N_o N_l \log(N_o)\log(N_l))$. For the
investigated device, only $\Niter^{FF}= 39$ GMRES iterations are necessary to solve
Eq.~\eqref{eq:sol_iter}, where $\Niter^{FF}$ is nearly independent of $N_o$ and $N_l$ (see also
\tabref{tab:GSM_materials}, second row). The black line in \figref{fig:convergence}(b) indicates
the runtime of the FMM, $t^{FF}_{FMM}$, which follows a $\mathcal{O}(N_o^3)$ dependence. This is
the expected behavior as the FMM involves finding the solution of an eigenproblem of size $N_o$.
The asymptotic advantage of the GSM becomes evident, e.g. at $N_l=200$, where for $N_o=1225$,
$t^{FF}_{FMM}$ becomes larger than $t^{FF}$.

For the computations at the SH, the GSM computational time still follows the $\mathcal{O}(N_o N_l
\log(N_o)\log(N_l))$ trend; however, as compared to the FF, approximately $3\times$ more
iterations, $\Niter^{SH} = 100$, are necessary to solve Eq.~\eqref{eq:sol_iter} at the SH. We
identify two possible reasons for this behavior: First, we found that for plane wave excitation,
the smaller the wavelength the more iterations were necessary for GMRES to solve
Eq.~\eqref{eq:sol_iter} and second, both the r.h.s. of this equation and its solution are more
inhomogeneous at the SH as compared to the FF.

The results for the generated SH, computed with $N_l=400$, are depicted in
\figref{fig:convergence}(c). Specifically, this figure shows the radiated power at the SH in the
direction of the incoming wave, normalized to the incident power at the FF. The data corresponding
to the solid line was obtained by using the modified inverse rule for the source polarization and
shows remarkably fast convergence to a value of $T^{SH}=1.163\cdot 10^{\minus 16}$ with increasing
$N_o$. By contrast, when using the unmodified rule $T^{SH}$ does not reach its asymptotic value
even for as many as $N_o=2209$ orders, despite the fact that the FF solution is already converged
if $N_o\gtrsim750$ [see \figref{fig:convergence}(a)]. Note that the old and correct factorizations
tend to slightly different values. Such a behavior has already been reported for linear
calculations in the FMM (\cite{l97josaa}).
\begin{table}
\caption{Characteristics of the nonlinear GSM for dielectric and lossy materials \label{tab:GSM_materials}}
\begin{center}
  \begin{tabular}{| c || c | c | c | c | c |}
            \hline
        %Rix & $\Niter_F$ & $\Niter_S$  & $T_F$ & $A_F$ & $T_S$ \\
        %\hline
    %1.45 & 21 & 38 & 0.972998 & $2.64\cdot 10^{\minus 8}$ & $2.88045\cdot 10^{\minus 17}$\\ \hline
    %2 & 39 & 100   & 0.917772 & $8.00\cdot 10^{\minus 7}$ & $1.16325\cdot 10^{\minus 17}$\\ \hline
        %3 & 105 & 400  & 0.825020 & $5.82\cdot 10^{\minus 6}$ & $4.57243\cdot 10^{\minus 18}$\\ \hline
        %0.3615 + 6.5005i & 105 & 400  & & &\\
        $n$ & $\Niter^{FF}$ & $\Niter^{SH}$  & $T^{FF}$ & $A^{FF}$ &\vspacer$T^{SH}$\vspacer\\  % ugly hack to get enough space for large superscripts
        \hline
    1.45 & 21 & 38 & 0.972998 & $2.6\cdot 10^{\minus 8}$ & \vspacer$2.88\cdot 10^{\minus 17}$\vspacer\\ \hline
    2 & 39 & 100   & 0.917772 & $8.0\cdot 10^{\minus 7}$ & \vspacer$1.16\cdot 10^{\minus 17}$\vspacer\\ \hline
        3 & 105 & 400  & 0.825020 & $5.8\cdot 10^{\minus 6}$ & \vspacer$4.57\cdot 10^{\minus 18}$\vspacer\\ \hline
                % $1.5\!+\!1i$ & 96 & 157  & 0.6201488 & 0.28341802 &  7.410311\cdot 10^{\minus 3}$ \\ \hline % at N=17, N_l=400
        $1.5\!+\!1i$ & 96 & 157  & 0.620149 & 0.283418 &  \vspacer$7.41\cdot 10^{\minus 19}$\vspacer \\ \hline % at N=17, N_l=400
                % $0.36\!+\!6.5i$ & 2160 & 1968  & 0.3811129 & 0.1932549 & $5.98785\cdot 10^{\minus 2}$ \\ % N=7, Nl=400, nonormals!
                $0.36\!+\!6.5i$ & 2160 & 1968  & 0.381113 & 0.193255 & \vspacer$5.99\cdot 10^{\minus 18}$\vspacer \\ % N=7, Nl=400, nonormals!
                % $0.36\!+\!6.5i$ & >4000 & >4000  & 0.4967839 & 0.13899913 & $6.076192\cdot 10^{\minus 2}$ \\ % N=7, Nl=400, nonormals!
                % $0.36\!+\!6.5i$ & $>4000$ & $>4000$  & 0.496784 & 0.13900 & $6.08\cdot 10^{\minus 18}$ \\ % N=7, Nl=400, nonormals!
        \hline
  \end{tabular}
\end{center}
\vskip -2em
\end{table}

Finally, we investigated the convergence properties of the GSM when it was applied to simulate
diffraction gratings made of three different lossless dielectric materials, a lossy dielectric,
and metallic components. The results, shown in \tabref{tab:GSM_materials}, provide valuable
insights into the characteristics of the GSM. Thus, clear trends can be observed in the case of
lossless dielectric gratings: the higher the refractive index $n$ is the more GMRES iterations
$\Niter^{FF}$ are necessary at the FF and the number of iterations $\Niter^{SH}$ at the SH
increases faster than $\Niter^{FF}$. The numerical absorption, $A^{FF}=1-T^{FF}-R^{FF}$, which in
the case of gratings made of lossless materials quantifies the degree to which the algorithm
conserves the energy, decreases with the index of refraction. The lossy material with $n=1.5+i$
can be efficiently analyzed by the GSM: the $\Niter^{FF}$ is relatively small and increases only
by 50\% for the calculation of the SH electromagnetic field.

The metallic grating requires a more detailed discussion. The derivation of the GSM formally
allows complex values of the permittivity $\varepsilon(\x)$ and the algorithm's implementation does
show that the method can be applied to simulate diffraction by lossy structures (see the $n=1.5+i$
example in \tabref{tab:GSM_materials}). However, we found out that the very large number of
required GMRES iterations makes the method particularly slow when applied to structures containing
very lossy components, namely a material with $n=0.36+6.5i$. We also found out that for metals
$\Niter$ increases strongly when $N_o$ increases (the presented results for $n=0.36+6.5i$ are for
$N_o=529$; increasing $N_o$ would require $\Niter>5000$ to achieve convergence of GMRES). Summing
up these results, the GSM in the presented form is most efficient and most suitable for shallow,
low refractive index contrast, dielectric devices. However, an improvement of the GSM using
curvilinear coordinates \cite{st13oe} allows one to efficiently calculate metallic structures as
well but we have not implemented this version of the algorithm in the nonlinear case.

%%%%%%%%%%%%%%%%%%%%%%%%%%%%%%%%%%%%%%%%%%%%%%%%%%%%%%%%%%%%%
\section{Application to SHG in textured slab waveguides}
\label{sec:Application} In this section we illustrate the versatility of our method by showing how
it can be applied to a problem of practical importance, namely to the maximization of the SHG in a
phase-matched configuration. Because of the weak nature of nonlinear optical interactions, the
radiated SH in practical settings is orders of magnitude smaller than the linear excitation.
Therefore field enhancing mechanisms are particularly important in this context as they can be
employed to increase the efficiency of SHG in practical applications. To this end, we consider the
SHG in a diffraction grating resonantly coupled to a slab waveguide, both optical devices being
made of quadratically nonlinear optical materials.

To add specificity to our problem, we assume that the nonlinear material in the device is GaAs
(refractive index $n = 3.4$), which due to its large optical nonlinearity is a particularly
suitable choice for our device application. It crystallizes in the zincblende lattice type, so
that it belongs to the crystal point group $\bar{4}3m$. Hence, the non-vanishing components of its
second-order susceptibility tensor are $\chi^{(2)}_{xyz} = \chi^{(2)}_{xyz} = \chi^{(2)}_{yxz} =
\chi^{(2)}_{yzx} = \chi^{(2)}_{zxy} = \chi^{(2)}_{zyx} =$~\SI{7.4e-10}{\meter\per\volt}
\cite{b08ap}, which is a relatively large value as compared to that of most other
non-centrosymmetric materials. As substrate we choose \CaF2 ($n_s = 1.4$) and consider air ($n_c =
1$) as cover, which provides the high refractive index contrast needed to achieve good field
confinement in the slab waveguide.

In order for the incident plane waves to couple and interact with the slab waveguide modes a
periodic optical grating that breaks the translational symmetry of the structure is placed on top
of the waveguide. As a result the propagation constant, $\beta_\nu(\lambda)$, of the $\nu$th slab
waveguide mode is folded within the first Brillouin zone of the grating. When the difference
between the tangential component of the incident wavevector, $\k_{\parallel}$, and the propagation
constant $\beta_\nu(\lambda)$ is equal to a multiple of the grating vector this waveguide mode can
be resonantly excited. In particular, one expects to observe strong field enhancement in the slab
waveguide at the corresponding resonance wavelengths and consequently significantly stronger
nonlinear optical interaction, an effect that has been recently used to demonstrate enhanced
optical absorption in such optical devices \cite{po07ol}.

\begin{figure}[b]
\begin{center}
        \includegraphics[width=0.49\textwidth]{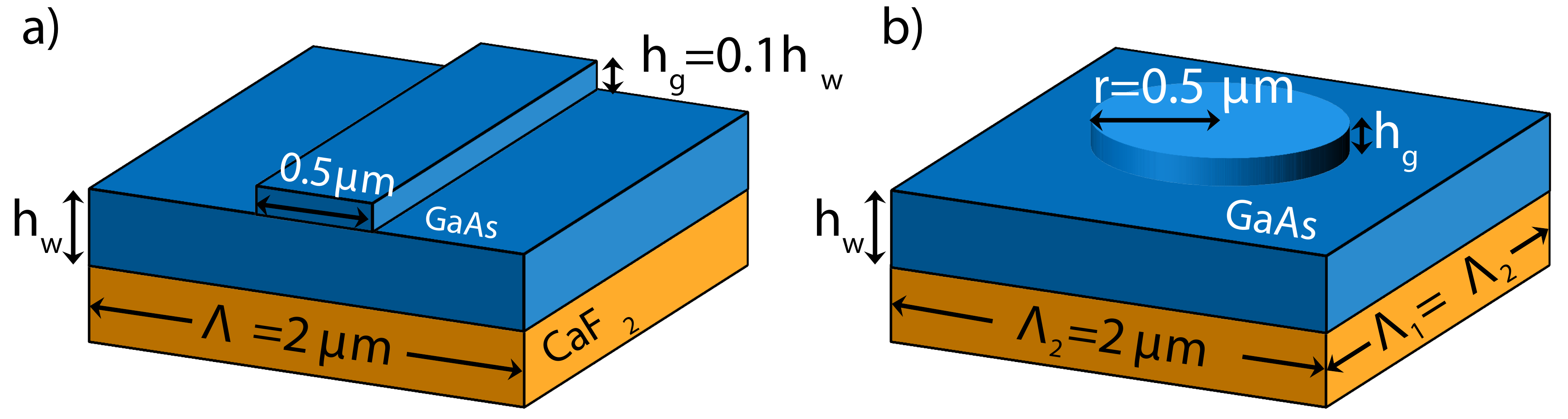}
\end{center}
\caption{\label{fig:SWG} a) Unit cell of a one-dimensional binary grating and b) two-dimensional cylindrical grating on top of a slab waveguide made of nonlinear GaAs on a \CaF2 substrate.
}
\end{figure}

In the context of SHG there are two kinds of resonances that can affect the nonlinear optical
response of the device: \textit{linear resonances} correspond to the grating-induced, resonant
excitation of a waveguide mode at the FF by a plane wave incident at a certain angle and
wavelength, $\lambda_{FF}$. We expect then that the optical field at the FF is strongly enhanced
and confined inside the waveguide, thus a stronger nonlinear polarization is created, which in
turn excites a stronger field at the SH wavelength, $\lambda_{SH}=\lambda_{FF}/2$. We call this
phenomenon of direct enhancement of the nonlinear response at the SH due to the excitation of a
linear resonance at FF an \textit{inherited resonance} at $\lambda_{SH}$. The second type of
nonlinear resonances are \textit{intrinsic resonances}, which are observed when a waveguide mode
is excited at the SH wavelength by the dipoles associated to the nonlinear polarization. In this
case the field at the SH is confined inside the waveguide and therefore shows remarkably large
radiated SH intensity. Note that the intrinsic resonances are not necessarily grating coupled with
the radiative continuum and therefore they can be viewed as nonlinear dark modes of the system
\cite{bp11n}. In particular, one expects that these resonances have large quality factor. We call
the inherited and intrinsic resonances \textit{nonlinear} resonances, since they are observed at
the SH. For the sake of simplicity, the modes causing the linear and nonlinear resonances are
denoted as linear and nonlinear modes. The SH wavelengths $\lambda_{SH}$ for which both types of
nonlinear modes are excited simultaneously are of particular interest because at these
wavelengths, this doubly resonant mechanism of SHG, can lead to a remarkably large nonlinear
optical response of the device. In the remaining of this section, these ideas are illustrated by
studying the nonlinear optical response of 1D and 2D implementations of these devices.

\subsection{One-dimensional binary gratings}
\label{sec:Application1D} We first investigate a 1D grating, which although having a simple structure,
provides us the necessary insights to completely understand the more complex 2D design. Thus, let
us consider a 1D binary GaAs grating on top of a GaAs slab waveguide, placed onto a \CaF2
substrate, see \figref{fig:SWG}(a). The height of the slab waveguide, $\height_w$, is a free design parameter and varies
from \SIrange{0.25}{0.5}{\micro\meter}. The period of the grating with filling factor $\rho=0.25$
is $\Lambda=$~\SI{2}{\micro\meter}, the height of the grating region is chosen in relation to the
height of the slab waveguide, $\height_g=0.1\height_w$.

At the FF, we consider a TM polarized plane wave with wavelength $\lambda_{FF}$ ranging from
\SIrange{3.6}{4.4}{\micro\meter} under $\theta=30^\circ$ incidence.
\uline{By considering the slab waveguide region as homogeneous region of the periodic grating, its nonlinearity is also calculated by the nonlinear GSM.}
For the calculation of the presented results $N_o=21$ diffraction orders and $N_l=138,\ldots,275$ GSM-layers were used, each with height of about \SI{2}{\nano\meter}.

\begin{figure}
    \centering
    \includegraphics[width=8cm]{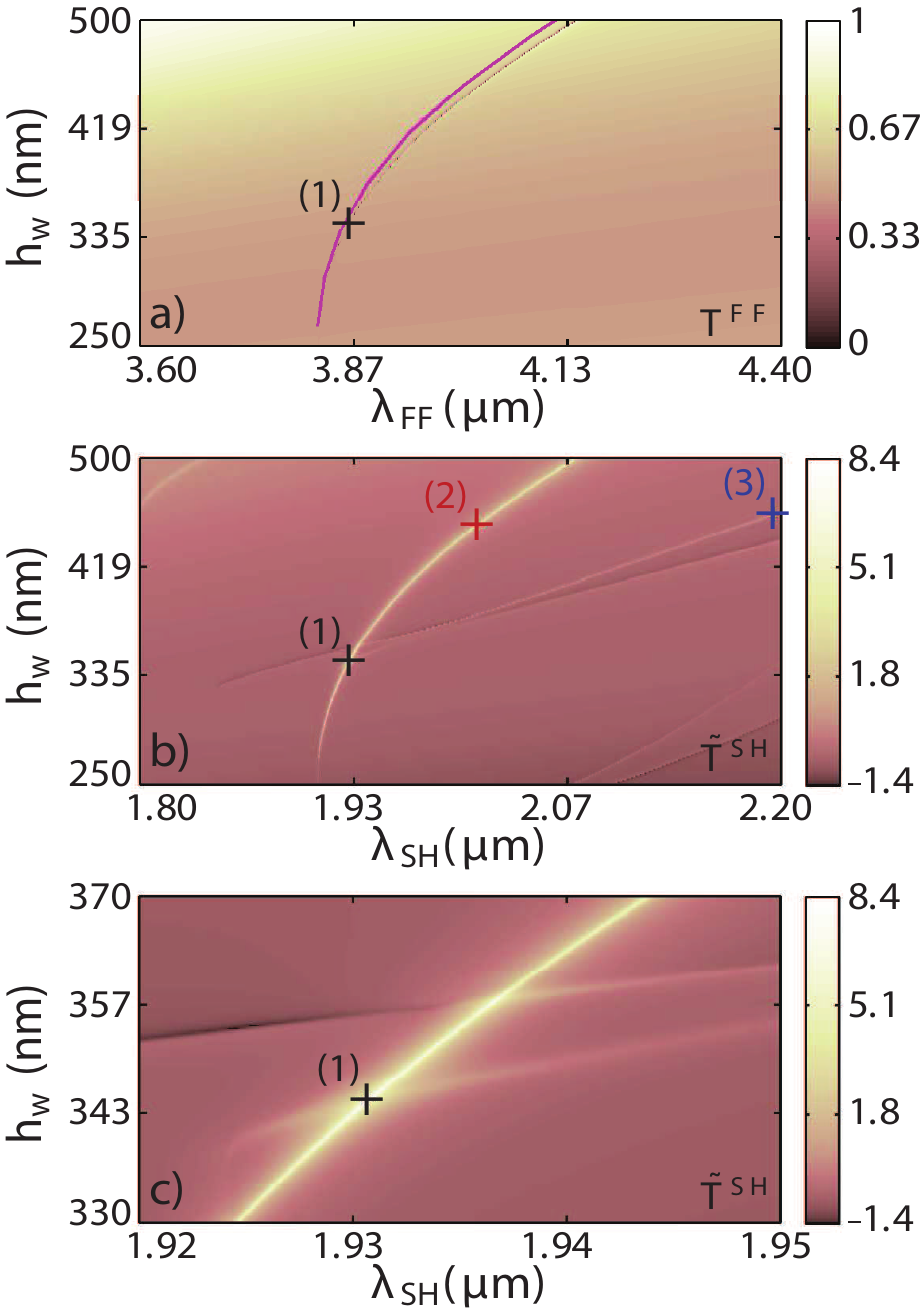}
\caption{Spectra of transmission coefficient for the binary texture determined for different
waveguide heights under normal incidence and for TM polarization. a) Spectra at the fundamental
wavelength (pink dispersion curve indicates analytical calculation of $\mathrm{TM}_0$ mode). b)
Spectra at second harmonic wavelengths (logarithmic scale). c) Zoom-in of the region around a
point where an intrinsic and inherited SH resonance dispersion curve cross.
\label{fig:SWM1D_Spectra} }
\end{figure}

\begin{figure}[t]
    \centering
    \centering
    \includegraphics[width=8cm]{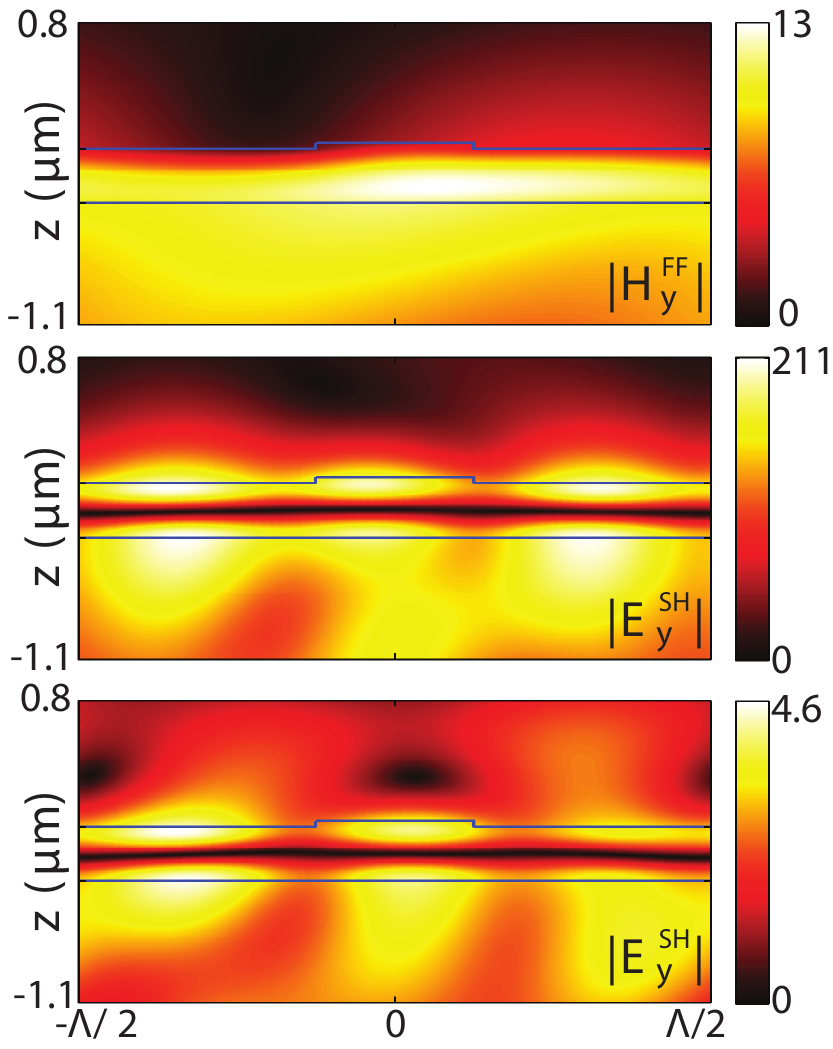}
\caption{Magnetic and electric field distributions at simultaneous resonance wavelength
$\lambda_{FF}=$~\SI{3.45}{\micro\meter} for $h_{w}=$~\SI{381}{\nano\meter}. a) $|H_y^{FF}|$ of the
fundamental $\mathrm{TM}_0$-resonance at the FF. b) $|E_y^{SH}|$ of simultaneous resonance at SH. c)
$|E_y^{SH}|$ of intrinsic $\mathrm{TE}_1$ mode at the SH wavelength
$\lambda_{SH}=\lambda_{FF}/2=$~\SI{1.725}{\micro\meter} for a TE polarized incident plane wave.
\label{fig:SWM1D_Fields} }
\end{figure}

Figure~\ref{fig:SWM1D_Spectra}(a) shows the transmitted intensity at the FF for a TM polarized
incident plane wave with wavelength $\lambda_{FF}$. The spectrum shows a smooth dependence on
$\lambda_{FF}$ and $\height_w$, except for a narrow stripe that corresponds to a sharp decrease of
the transmission. This change in transmission can clearly be attributed to the excitation of a
$\mathrm{TM}_0$ waveguide mode. Thus, an analytically calculated resonance wavelength
\cite{wgp14spie}, which is represented by the pink line in the spectrum in
\figref{fig:SWM1D_Spectra}(a), is in close proximity to the band of reduced transmission seen in
the spectrum. The analytical calculation is obtained for a infinitesimally small perturbation of
the slab waveguide, whence the agreement is nearly perfect for low $\height_w$ and slowly
deteriorates as $\height_w$ and therefore $\height_g$ increases. Moreover,
\figref{fig:SWM1D_Fields}(a) shows the $y$-component of the magnetic field at the FF, $H_y^{FF}$,
inside and around the textured waveguide, which is indicated by blue lines, corresponding to the
black cross (1) in \figref{fig:SWM1D_Spectra}(a), i.e. to $\lambda_{FF}=$~\SI{3.861}{\micro\meter} and
$h_{w}=$~\SI{345}{\nano\meter}. This field profile clearly matches that of the $\mathrm{TM}_0$
mode. Note that for a 1D grating under non-conical incidence, the TE and TM components of the
electromagnetic field decouple. Accordingly, there is only a single TM mode excited under
TM-incidence, in the considered $\height_w$-$\lambda_{FF}$ domain.

In order to better visualize the SHG enhancement, we plot in \figref{fig:SWM1D_Spectra}(b) the
logarithm $\tilde T^{SH}=\log_{10} T^{SH}$ of the SH intensity, emitted in the direction of the
incident wave, for the same incident fundamental field as in \figref{fig:SWM1D_Spectra}(a). The
transmission coefficient, $T^{SH}$, is normalized to the SHG in a bulk layer of GaAs having the
same volume as the textured grating. This figure clearly shows an increase of the SH intensity due
to the excitation of the inherited TM modes and intrinsic TE modes. No intrinsic TM modes are
excited in this case. This is due to the particular symmetry properties of $\bfchi$: for a
TM polarized fundamental field $\E(\x) = (E_x(\x),0,E_z(\x))$, only the $y$-component (i.e. the TE
component) of the nonlinear polarization is non-vanishing.

At $\lambda_{SH}=$~\SI{2.196}{\micro\meter} and $h_{w}=$~\SI{460}{\nano\meter}, corresponding to
the blue cross (3) in \figref{fig:SWM1D_Spectra}(b), one can achieve an intensity enhancement of
$T^{SH}=10^{0.55}\approx 3$ due to the excitation of an intrinsic $\mathrm{TE}_1$ mode. Although
the intensity is only three times higher than in the case of the bulk material, it is still
significantly larger than the surrounding designs of textured waveguides. The excitation of the
inherited mode yields much stronger enhancement of SHG, namely at
$\lambda_{SH}=$~\SI{2.011}{\micro\meter} and $h_{w}=$~\SI{451}{\nano\meter} (red cross (2)),
$T^{SH}=10^{4.78}$.

The strongest SH enhancement of $T^{SH}=10^{8.5}$ is achieved for simultaneous excitation of the
inherited $\mathrm{TM}_0$ mode and intrinsic $\mathrm{TE}_1$ mode at
$\lambda_{SH}=$~\SI{1.931}{\micro\meter} and $h_{w}=$~\SI{345}{\nano\meter} (black cross (1)). A
zoom-in of the region of the doubly resonant excitation is presented in
\figref{fig:SWM1D_Spectra}(c). The electric field profile $E_y^{SH}$ at the SH for this particular
configuration, plotted in \figref{fig:SWM1D_Fields}(b), shows high field enhancement inside the
waveguide-grating region with two pronounced maxima at the top and bottom of the structure. This
field profile can be explained by inspecting the distribution of the field $E_y^{SH}$ of the
intrinsic $\mathrm{TE}_1$ mode at $\lambda_{SH}=$~\SI{1.725}{\micro\meter} and
$h_{w}=$~\SI{381}{\nano\meter}. Thus, for TE-polarized incident waves with
$\lambda_{FF}=2\lambda_{SH}=$~\SI{3.45}{\micro\meter}, the electric field around the textured
waveguide closely resembles the profile of a $\mathrm{TE}_1$ mode, as can be seen in
\figref{fig:SWM1D_Fields}(c). However, the field enhancement is not as strong as in the case of
the linear $\mathrm{TM}_0$ mode, as per \figref{fig:SWM1D_Fields}(b).

This analysis fully explains the dominant feature of the SH response of the 1D device when the two
types of resonances are simultaneously excited, namely the overall SH intensity is mainly
determined by the strong enhancement inside the waveguide of the fundamental field due to the
linear resonance. The field profile, however, is determined by the profile of the intrinsic
$\mathrm{TE}_1$ mode.

\subsection{Two-dimensional cylindrical gratings}
\label{sec:Application2D} The ideas presented in the previous section can be easily extended to a
similar 2D device. To illustrate this, we consider the GaAs grating in \figref{fig:SWG}(b) with unit cell consisting of a cylinder with radius, $r=$~\SI{500}{\nano\meter} and relative height $\height_g=0.1\height_w$,
placed at the center of a square with side-length $\Lambda=$~\SI{2}{\micro\meter}. The GaAs
waveguide is placed on a \CaF2 substrate and its height, $\height_w$, varies from
\SIrange{660}{802}{\nano\meter}.
For the presented computationally expensive parameter sweep, a total of $N_o=81$ diffraction
orders and $N_l=350,\ldots,450$ GSM-layers with individual height of \SI{2}{\nano\meter} were used.
\uline{By checking the convergence with respect to $N_o$ for a fixed }$h_w=$~\SI{700}{\nano\meter}  \uline{and varying wavelength, we found that $N_o=81$ already yields qualitatively correct results.}

\begin{figure}[b]
    \centering
    \includegraphics[width=8cm]{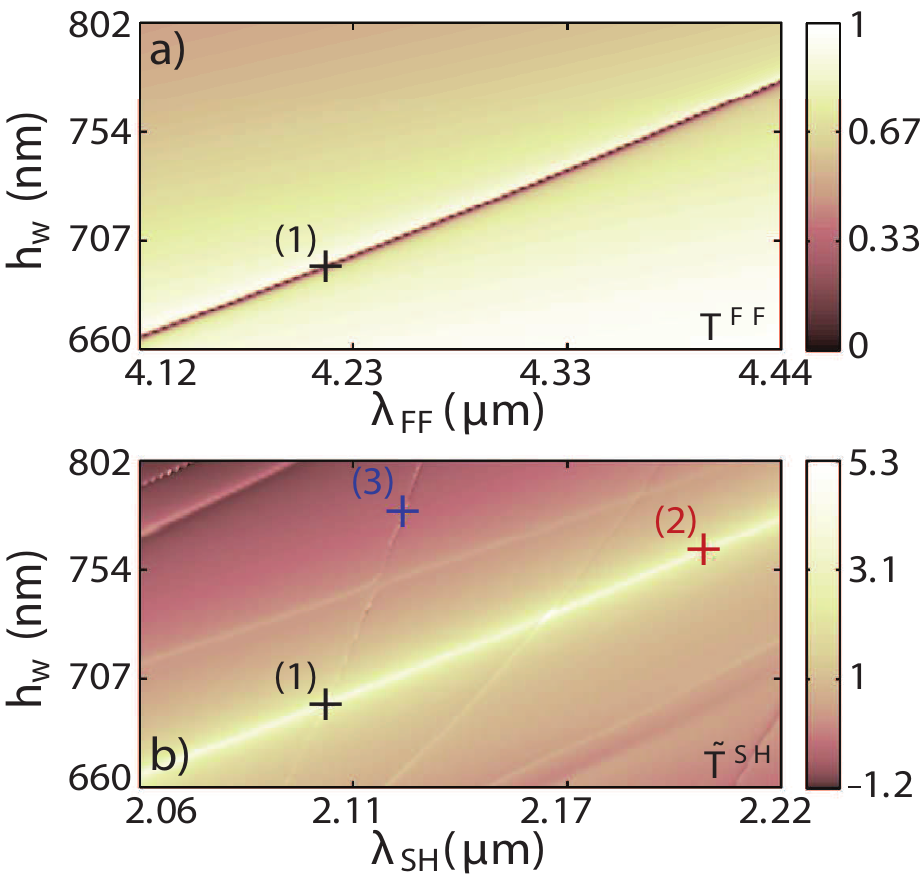}
\caption{Spectra of transmission coefficient for the cylindrical texture determined for different
waveguide heights under normal incidence. a) Spectra at the fundamental wavelength. b) Spectra at
SH wavelengths (logarithmic scale). \label{fig:SWM2D_Spectra} }
\end{figure}

The results for normally incident, TM-polarized plane waves with wavelength
$\lambda_{FF}=$~\SIrange{4.12}{4.44}{\micro\meter} are summarized in \figref{fig:SWM2D_Spectra}.
As in the 1D-case, the transmission spectrum at the FF reveals the resonant excitation of a
$\mathrm{TM}_0$ mode, seen in \figref{fig:SWM2D_Spectra}(a) as a dark stripe.

The spectrum of radiated SH in the direction of the incoming wave, presented in
\figref{fig:SWM2D_Spectra}(b), exhibits distinct maxima along the dispersion curves of the
inherited and intrinsic resonances. For example, for the inherited $\mathrm{TE}_0$-resonance at
$\lambda_{SH}=$~\SI{2.126}{\micro\meter} and $\height_w=$~\SI{780}{\nano\meter} (blue cross (3)) the
SHG enhancement as compared to a uniform GaAs layer with the same volume is
$T^{SH}=10^{0.77}\approx5.9$. The inherited resonance at $\lambda_{SH}=$~\SI{2.201}{\micro\meter} and
$\height_w=$~\SI{763}{\nano\meter} (red cross (2)) yields $T^{SH}=10^{3.95}$.
\begin{figure}
    \centering
    \includegraphics[width=8cm]{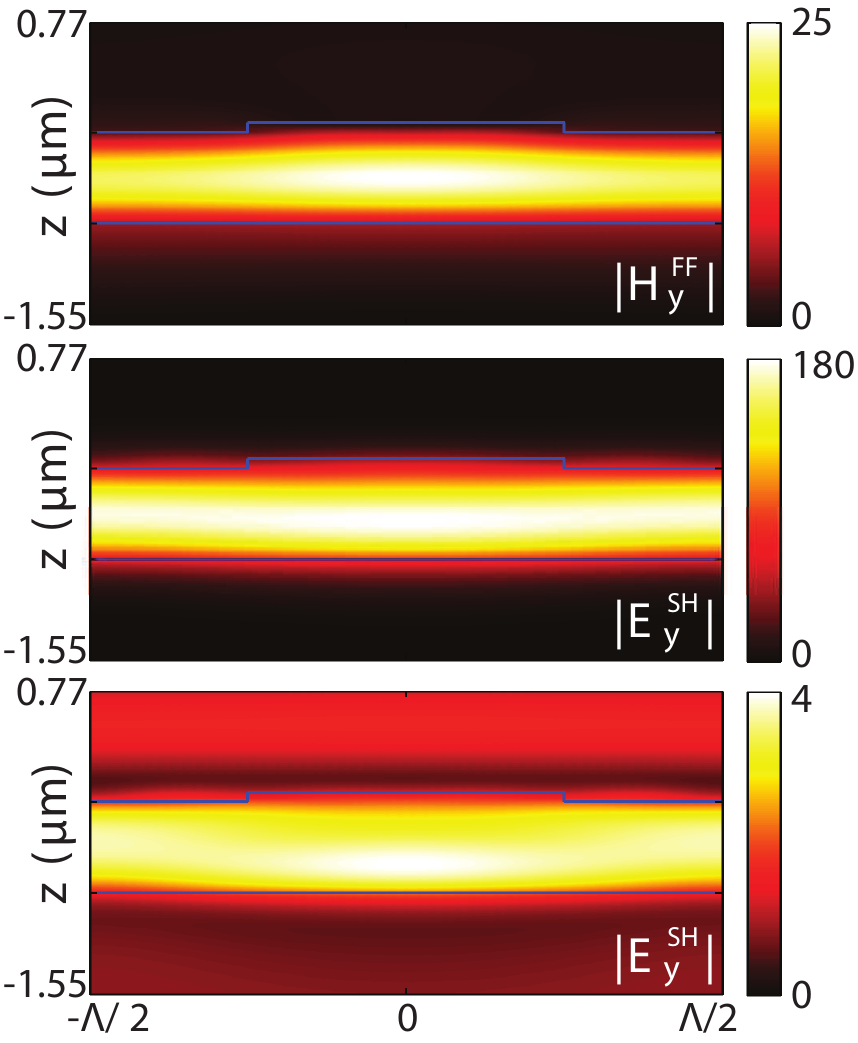}
\caption{Magnetic and electric field distributions at simultaneous resonance wavelength
$\lambda_{FF}=$~\SI{4.213}{\micro\meter} and $h_{w}=$~\SI{696}{\nano\meter}. a) $|H_y^{FF}|$ of the
fundamental $\mathrm{TM}_0$-resonance at the FF. b) $|E_y^{SH}|$ of simultaneous resonance at SH. c)
$|E_y^{SH}|$ of intrinsic $\mathrm{TE}_0$ mode at SH wavelength
$\lambda_{SH}=\lambda_{FF}/2=$~\SI{2.107}{\micro\meter} for TM polarized incident plane wave.
\label{fig:SWM2D_Fields} }
\end{figure}

We have designed a 2D periodic device in which inherited and intrinsic resonances can be simultaneously
excited, the corresponding parameters being $\lambda_{FF}=$~\SI{4.213}{\micro\meter} and
$\height_w=$~\SI{696}{\nano\meter} (black cross (1)). This double resonance leads to an intensity
enhancement of $I^{SH}\approx 10^5$ as compared to the reference uniform layer. The spatial
profile of the magnetic field, calculated in the $x\minus z$-plane passing through the center of
the cylinder ($y=0$) is displayed in \figref{fig:SWM2D_Fields}(a). The field distribution closely
resembles the profile of the $\mathrm{TM}_0$-mode, showing strong field enhancement and increased
field confined in the waveguide as compared to the 1D-case. Our simulations suggest that normal
incidence and a larger waveguide thickness favor stronger confinement. Moreover, the intrinsic
$\mathrm{TE}_0$ mode profile is given in \figref{fig:SWM2D_Fields}(c). As in the 1D-case, it
strongly influences the shape of the SH-field inside the textured waveguide, as per
\figref{fig:SWM2D_Fields}(b).

%%%%%%%%%%%%%%%%%%%%%%%%%%%%%%%%%%%%%%%%%%%%%%%%%%%%%%%%%%%%%
\section{Conclusion}
\label{sec:Conclusion} We have presented an extension of the generalized source method to analyze
SHG in 2D periodic structures containing non-centrosymmetric, quadratically nonlinear optical
materials. The linear and nonlinear parts of the calculations are decoupled in the undepleted pump
approximation and a three-step computational process was derived. We introduced the correct
Fourier-factorization rule for inhomogeneous problems and its beneficial effect on the convergence
in the nonlinear calculations was clearly shown. In a benchmark structure the asymptotic runtime
estimate of ${\mathcal O}(N\log(N) N_l\log(N_l))$ was validated, which renders the GSM especially
suitable for modeling low permittivity contrast, dielectric gratings.

As a practical application we optimized the design of 1D and 2D textured GaAs slab waveguides to
achieve maximal radiated SH by simultaneous excitation of linear and nonlinear slab waveguide
modes. In a addition to an analytical estimate for the resonance frequencies of the textured slab
waveguide, investigation of the field profiles allowed us to identify the nature of the resonances
responsible for the remarkably high intensity enhancement of more than $8$ orders of magnitude in
comparison to the SH radiated by a uniform slab. Importantly, the formalism presented in this
study can be extended to several other key nonlinear optical processes, including surface SHG from
centrosymmetric materials and higher-harmonic generation. Our method can be further extended
beyond the undepleted pump approximation by allowing a self-consistent coupling between the FF and
SH optical waves. In a similar manner, Kerr-nonlinear optical materials can be considered as well,
a topic that we plan to address in a future study.

% Things to mention: anisotropic materials
% easily extendable to other non-linear processes which can be decoupled using the undepleted pump approximation: nonlocal bulk, sum/difference-frequency generation and surface shg
% remove undepleted pump approximation
% kerr nonlinearity.
% Comparison to experiments

% efficiency dependent on material, asymptotic runtime actually $O(N\log(N) N_l N_{iter}$. $N_l$ and $N$ not independent for given accuracy.\\
% - Metals particularly difficult due to high field enhancement. high condition number of problem\\
% - Watanabe stuff

%%%%%%%%%%%%%%%%%%%%%%%%%%%%%%%%%%%%%%%%%%%%%%%%%%%%%%%%%%%%%
\acknowledgments The work of M. W. was supported through a UCL Impact Award graduate studentship
funded by UCL and Photon Design Ltd and by the Engineering and Physical Sciences Research Council,
grant No EP/J018473/1. The authors acknowledge the use of the UCL Legion High Performance
Computing Facility (Legion@UCL) and associated support services in the completion of this work. N.
C. P. wishes to acknowledge the hospitality of Y. S. Kivshar and the Nonlinear Physics Centre of
the Australian National University during a visit when this paper was being written.

%\bibliographystyle{osa}
%\bibliography{main}

\end{document}